\documentclass[runningheads]{llncs}
\usepackage[title]{appendix}
\usepackage[ruled, vlined, linesnumbered]{algorithm2e}
\usepackage{multirow}
\usepackage{subcaption}
\usepackage{xcolor}

\newcommand{\etal}{\textit{et al}. }

\usepackage{graphicx}
\begin{document}
%

\title{User profiling using smartphone network traffic analysis}
\titlerunning{}
%


\author{Ayush Bahuguna\inst{1} \and Ashutosh Bhatia\inst{1} \and Kamlesh Tiwari\inst{1} \and Deepak Vishwakarma\inst{2}}
\authorrunning{Ayush Bahuguna  et. al.}
%

\institute{ Dept. of  Computer Science and  Information Systems \\ Birla Institute of Technology and Science, Pilani, Rajasthan 333031, India \\ 
\{f2015144, ashutosh.bhatia, kamlesh.tiwari\}@pilani.bits-pilani.ac.in
\and 
  Center for Artificial Intelligence and Robotics, DRDO, Bangalore \\  
 vishwakarma.deepak@gmail.com }

\maketitle              
\begin{abstract}

The recent decade has witnessed phenomenal growth in communication technology. Development of user-friendly software platforms, such as Facebook, WhatsApp etc. have facilitated ease of communication and thereby people have started freely sharing messages and multimedia over the Internet. Further, there is a shift in trends with services being accessed from smartphones over personal computers. To protect the security and privacy of the smartphone users, most of the applications use encryption that encapsulates communications over the Internet. However, research has shown that the statistical information present in a traffic can be used to identify the application, and further, the activity performed by the user inside that application.  In this paper, we extend the scope of analysis by proposing a learning framework to leverage application and activity data to profile smartphone users in terms of their gender, profession age group etc. This will greatly help the authoritative agencies to conduct their investigations related to national security and other purposes. 

\keywords{Encrypted Traffic Analysis, User Profiling, Deep Learning}
\end{abstract}

\section{Introduction}
In the era of information and communication technology, to ensure national security, maintain law and order and to perform investigations, various agencies of the government require  access of the digital data of their citizens.  To facilitate this, the Ministry of External Affairs, Government of India on Dec. 20, 2018 has expanded the scope of IT Act 2000 \cite{it2000} in order to make various government agencies capable to monitor and intercept the Internet traffic belonging to Indian citizen under their jurisdiction. On one hand, such a decision would empower various government agencies to perform the investigation. On the other hand, it puts  privacy of the citizens at risk to compromise. Ensuring national security while maintaining the privacy of a citizen is a difficult task. 

Many application service providers on the Internet, use end-to-end encryption to protect the data privacy of their users. For example, messages sent on WhatsApp are protected by end-to-end encryption which makes the data intangible for a third party trying to intercept these messages somewhere over the Internet. So, even when a networking device or access point gets compromised, utilizing the captured encrypted data for various practical or nefarious purposes becomes a challenging task. 

Concomitant research in this area has probed the possibility of identifying a few particular applications or activity/action performed by the user in that application. Nguyen \etal \cite{surv1} describe an in-depth survey evaluating different methods for encrypted mobile traffic classification. Aceto \etal \cite{surv2} classify analytical frameworks specifically for smartphones. This paper also specifically focuses on the learning frameworks for mobile devices. Much work in literature has been done by clustering packets as flows where multiple bursts of packets defined as a set of packets occurring withing a burst threshold, are then aggregated into clusters based on IP and port mappings. Random forests have predominantly proven as a useful variant of machine learning classifiers in the context of flow classification, owing to their robustness to noise and low computational requirements during testing conditions \cite{rf1,rf2,rf3,rf4}. Conti \etal \cite{rf3}, use agglomerative hierarchical clustering in conjunction with dynamic time warping to merge clusters of packets in a manner that maximises prediction confidence for a random forest classifier generating activity labels. Hidden Markov models present an approach modeling future states of a stochastic system based on observed or {\it hidden} state variables. This sets the framework for predictive sequential algorithms which have been proven robust \cite{hmm0,hmm1,hmm2} at modeling features like prices, volatility and regime classification. Fu \etal \cite{ensemble1} describe a framework that labels traffic based on usage type (such as text, multimedia, location sharing) using an ensemble classifier model where a base classifier proposed as a random forest is boosted with outlier detection using combined \textit{k}-means and Hidden Markov models. Auld \etal \cite{bnn} apply Bayesian neural networks, a modification of the standard ANN model where Bayesian posterior calculations is applied for optimisation of network weights, for application classification of flows. An entropic confidence scheme is used to selectively pick model predictions that increase overall model accuracy.

Knowing the atomic activity performed by a user is not sufficient to profile him in terms of his/her social or professional conducts.  The kind of question a particular investigating agency might be interested are: What is the gender of the user? What is the age group? Whether the user is a student or professional? And much more. These questions correspond to profiling a user. It is tough, to find labeled data for such supervised learning tasks due to privacy issues. At the same time it is difficult to determining distinguishing features that are accurate and mathematically meaningful for such anthropomorphic attributes. However, there is a need to develop systems that can provide answers to these questions. 
This paper proposes a framework to perform user profiling by applying deep learning techniques on the captured encrypted traffic of the mobile users. In particular, the proposed framework first tries to identify the mobile application to which a captured traffic belongs, then exact activity performed by the user in that application, and finally, it uses this information to identify the gender, professional activity and the age group of the mobile user. To best of our knowledge, this is the first work of it's kind, attempting user profiling over the encrypted traffic data.

\section{Background}
A natural approach to analyse network traffic will underlie selecting mappings from input to feature spaces and then to apply suitable ML models. The first hurdle which this approach faces is that, the entire model suffers if feature selection does not identify viable artifacts from the underlying network traffic data. Another issue that the data has temporal nature. So majority of machine learning models used for the problems find it difficult to decide the length of temporal dependence among packets getting accumulated. While considering network traffic data alone, it is not known when a particular activity begins or ends on a targeted device, hence features without time-scale invariance or convergence are not immediately informative and lose information on normalization.

In this respect, deep learning techniques specialised in isolating mappings for input spaces that maximise variance across classes. Model variables are optimised by means of gradient descent techniques to map data to outputs. Research in deep learning and artificial neural networks has covered major ground in the field of automating complicated control systems, however in the fields of networking analysis, these models face significant competition from statistical or stochastic contemporaries. Often, payoff with deep models is achieved only after duly adapting the \textit{model} to the data, and not just the other way round by preprocessing features. Recurrent neural networks (RNNs) are a well-defined subset of artificial neural networks devised as an optimisation for sequential data, and will be a major focus as detailed in section \ref{methods}. 

A common challenge in the domain of deep learning is to devise a general guideline for structuring models based on the underlying problems in question. Recent developments in artificial neural network architectures have shown that models that have been optimised to capture meaningful variance in particular output spaces may hold a better chance at learning different output spaces that seem homogeneous based on surface level statistics. Under the framework of transfer learning \cite{transfer}, it is possible to develop useful models by retraining output layers of another model trained with some degree of success on a different problem, that can ultimately achieve a similar degree of success if the problem formats are similar enough and the format of data to be processed by the network does not majorly change. Application and activity labels, which relate well with underlying semantics in data, might possibly translate a priori information well to anthropomorphic behavior and hence transfer learning or ensemble classification might be feasible.

\subsubsection{Long short-term memory}
Recently, great interest has been directed towards recurrent neural networks (RNNs), a reinterpretation of the standard artificial neural network architecture, in which input data are fed to sequential neural blocks, predicting an output sequence at each cell step. Since these blocks represent a recursive circuit, these networks take advantage of an effective artificial ``memory", similar in essence to how sequential circuits provide computational memory, allowing specialized networks that can track correlations in temporal data accurately.

The long short-term memory (LSTM) model was a reinterpretation of RNNs in response to criticism of RNNs for poor retention of features with increasing number of steps for which the model is unrolled. LSTM models employ sequential neural gates, where a gate $g$ is a data structure whose output $g_t$ at an instance $t$ along the input sequence is determined by the equation 
\begin{equation}
p_t = \sigma_{g}(W_{g}x_t + U_{g} c_{t-1}+b_g),
\end{equation}
for inputs $x_t$ and an entry-wise activation function $\sigma_{g}$ (the \emph{peephole} LSTM implementation \cite{peep} where the cell state $c_{t}$ is visible to gates). The trainable parameters are the matrices $W$ and $U$ as weights between current input vectors and previous cell state and the gate output, and a bias vector $b$ added to the gate's raw input. A LSTM cell employs three such gates, each with a sigmoid activation. The forget and input neural gates $f$ and $i$ respectively drive the current cell state by the update equation 
\begin{equation}
c_{t}=f_{t} \circ c_{t-1} + i_{t} \circ \sigma_{c}(W_{c}x_{t} + U_{c}c_{t-1}+b{c}), 
\end{equation}
where $\circ$ is the Hadamard product operator and $\sigma_{c}$ is the hyperbolic tangent function. The final output gate $o$ is used to generate a prediction $\theta$ by the equation 

\begin{equation}\label{eq:cell}
\theta_{t+1} = o_{t} \circ \sigma_{h}(c_{t}),
\end{equation}

\noindent where for our model, $\sigma_h = \sigma_{c}$. As their names suggest, the gates help to achieve obsolescence, modification and propagation of cell features and hence more selective paramatrizations of temporal features. Simpler procedures such as gated recurrent unit (GRU) models assume lower update function complexity and lesser gates, achieving similar predictive results.


\section{Proposed Approach} 
\label{methods}

\subsection{Ground Truth Construction}
In this section we describe our traffic generation test-bed that emulates data captured over an authoritative/compromised access point serving a targeted device. Behaviour and trait identification are fairly distinct avenues for analysis, when considering what labels they expect for their data. Behavioural analysis is more focused on identifying applications and underlying implementations since these predominantly control the actions that the user can perform when interacting with their device; trait identification simply focuses on clustering users based on relevant user characteristics. This is the discrepancy that our work aims to resolve, and in this section we describe our approach to model a dataset that can facilitate this.

\begin{table}[t]
\centering

\caption{Structure of the dataset.}
\resizebox{\textwidth}{!}{%
\begin{tabular}{|c|c|c|c|c|c|c|c|c|}
\hline
Application                    & \multicolumn{2}{c|}{Facebook} & \multicolumn{2}{c|}{YouTube} & \multicolumn{2}{c|}{WhatsApp} & Gmail               & \multirow{2}{*}{Impertinent} \\ \cline{1-8}
Action                         & post\_text    & post\_image   & play\_video     & comment    & send\_message  & send\_image  & mail                &                              \\ \hline
\multirow{2}{*}{No. of series} & 237           & 423           & 102             & 129        & 83             & 203          & \multirow{2}{*}{81} & \multirow{2}{*}{7068}        \\ \cline{2-7}
                               & \multicolumn{2}{c|}{660}      & \multicolumn{2}{c|}{231}     & \multicolumn{2}{c|}{286}      &                     &                              \\ \hline
\end{tabular}%
}
\label{dataset}
\end{table}

An automated approach was pursued to generate and capture network traffic with associated application and action labels, for which the open-source automation platform Appium \cite{appium} was used. The tool is built upon the Selenium WebDriver \cite{selenium} which provides various client drivers for automating interactions in Android, iOS or HTML environments. In Appium, the creation and handling of WebDriver sessions on devices hosting the automation is managed using a Node.js webserver that communicates automation commands to implementation-specific listeners spawned on the device. Scripts were written for a fixed set of timestamped actions for four mainstream Android applications - Facebook, Gmail, WhatsApp and YouTube - and individually run using an iterative testing framework that sniffs all communications simultaneously using tshark. Timestamps were then used to isolate streams which are then filtered during the labeling process. Streams for Facebook, WhatsApp and YouTube each had two possible activity labels. Table \ref{dataset} shows the structure of our proposed dataset. Only 75 automated actions were performed per action class. The discrepancy in volume of impertinent streams even in a heavily controlled test-bed and the variation in the number of streams servicing an activity are easily distinguishable.  

Although applications may depend entirely on third party services, most popular contemporary applications communicate to dedicated servers and such information can be used on a surface level when filtering packets. For most applications, packets can be differentiated using hostnames shared in Server Name Indication (SNI) fields of SSL communications. On the contrary, for applications with end-to-end encryption like WhatsApp, DNS queries in the captured traffic are required for identifying TCP traffic using the query response address \cite{wae}. In that case, a DNS query needs to be forced in the beginning of traffic capture, to ensure no packets are missed. For this reason, WhatsApp communications rarely require SSL encryption except when exchanging encryption secrets.

For the task of user profiling, 
we have used an Android application tPacketCapture \cite{tpc} for collecting data without application labels, solely for the purpose of identifying user profiles. The application monitors interfaces for a virtual private network (VPN) using the Android \texttt{VPNService} API through which all the communications for the device are routed. Importantly, all Android enforces unique identifiers for applications and this allows the service to provide a handy filtering mechanism for selective logging, which makes this approach viable for generating data from volunteers. Without an intrusive active learning approach, however, such a procedure cannot be expected to capture variance on an atomic level for individual activities.

\subsection{Preprocessing} 
All communication packets  generated as above have to be processed into a format compatible with ML models. 
Most deep learning methodologies impose different normalization requirements from data at this stage. We use a sign bit $s_t$ and a positive floating point value $p_t$ that linearly maps packet sizes upto a threshold of 2048 to values in the range $(0,1]$. Differences between timestamps, $T_t$ - $T_{t-1}$ are also similarly mapped to a floating point value $delay_t$ using a threshold for delay. This gives us a normalized feature vector $<\sqrt{delay_t}, s_t, \sqrt{p_t}>$ for each packet in a stream. The square root is calculated to increase the variance between features with low values of packet sizes or delays, as these can well be expected to represent the largest portion of captured data.

This format of preprocessing  data was influenced by the way recurrent neural networks extract features from inputs.
Our implementation was done using TensorFlow's \texttt{LSTMCell} class. The \texttt{tensorflow.nn.dynamic\_rnn} method allows training over batches of sequence-label pairs, however the framework imposes limitations over the number of steps for which the cell is unrolled over a batch of several sequences. This has the advantage that it fixes the scales of sequence length in which the model needs to learn patterns, and facilitates accelerated batch-normalising of loss gradients at different levels of the network. Every sequence is hence clipped or padded to a length of 2048 entries and we apply pooling with an exponential gradient in steps of 16 entries: the first two batches of 16 entries are considered as is, the next 32 are shortened to 16 by applying a pooling function with a step of 2, the next 64 with a step of 4 and so on until the last 1024 entries. This provides us with a series of 128 vectors for every stream. Both max-pooling and average pooling are used, so the first 32 are entries are appended with the previous entry in the series as a redundant input to accordingly fix the input size of each entry at six.

\subsection{Classification}
With the format of described dataset, we aim for models that can responsibly provide granular labels for user activity. In this section we describe modular classifiers developed for successive application and in-app activity.

\begin{figure}[t]
    \centering
    \includegraphics[width=.7\columnwidth, height=.55\columnwidth]{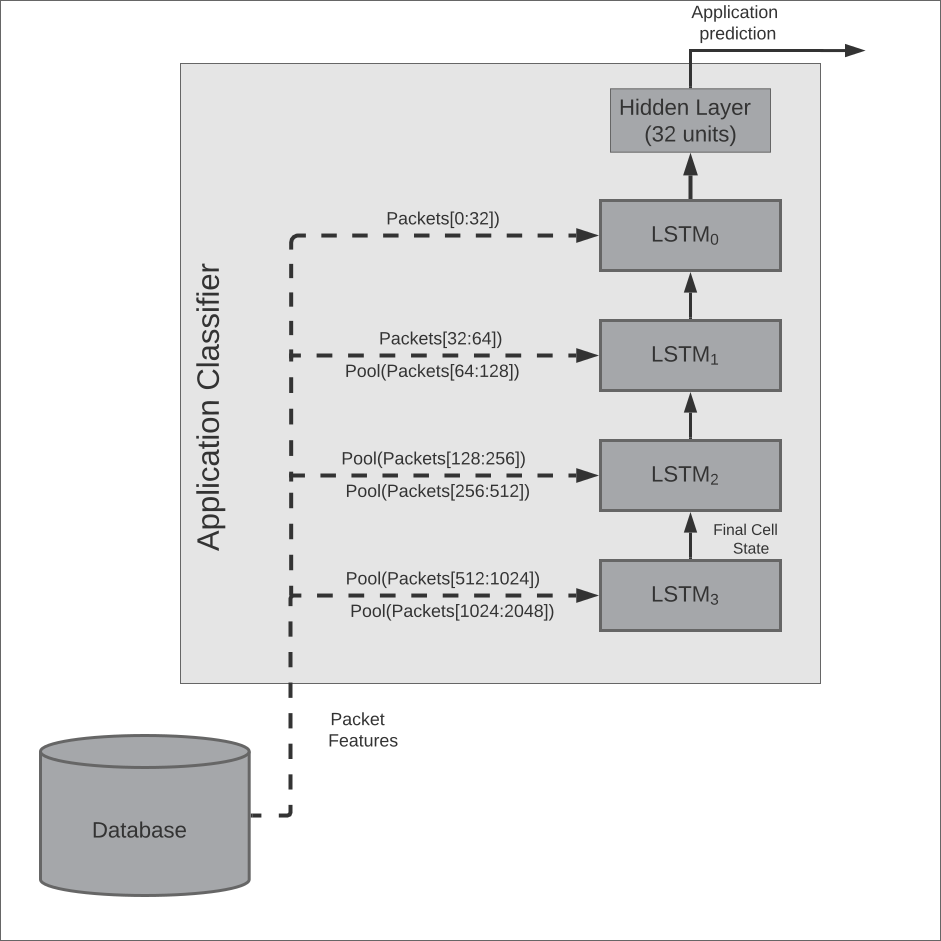}
    \caption{Classifier for application labels. The first three cells generate features at different scale changes along the exponential pooling process.}
    \label{fig:baselstm}
\end{figure}

\subsubsection{Application classifier}

Fig. \ref{fig:baselstm} describes the layout of a multiple-RNN analysis model that generates stream-wise application labels. An ensemble of stacked LSTM cells with 32 units are used to read packet features, where the first three LSTMs cells are exposed to pooled features at different scales to learn large scale features, and a final LSTM cell is tasked with processing packet-level features for the first 32 packets. These features are processed by an output layer to give application predictions. For larger stream sizes, features for trailing packets look homogeneous between different applications and the first $n$ packets usually capture sufficient variance \cite{firstn}. This explains the reversed processing of the time series, as the cell exposed to the most discriminating features should not be placed earlier in a recurrent learning pipeline in order to suppress information loss. 
\subsubsection{Activity classifier}
The application classifier passes its output as well as it's cell state to secondary classifiers for activity labeling. These models were implemented as single LSTM cells take the original time series as well as application classifier output and cell states. Since the cell is placed further in the model pipeline and preceding cell states are visible, only one LSTM cell is used except with variable neural gate depth. Thresholded application predictions decide which activity classifier to drive, however the correct application classifier is optimised during training independent of application prediction. Usually in training multiple classifier models, preceding models need to be trained to completion before successor models can be trained to completion. Parallel training for modular neural networks is however feasible and is exploited in training schemes which use multiple losses. Since we have separate labels for application and activity, there are equivalently different loss functions for the two and hence there is the choice of designing independent classifiers or allowing back propagation of loss gradients between them. Preliminary tests however suggested that the models fare better when learning independently in parallel.

\begin{figure}[t]
    \centering
    \includegraphics[width=.7\columnwidth]{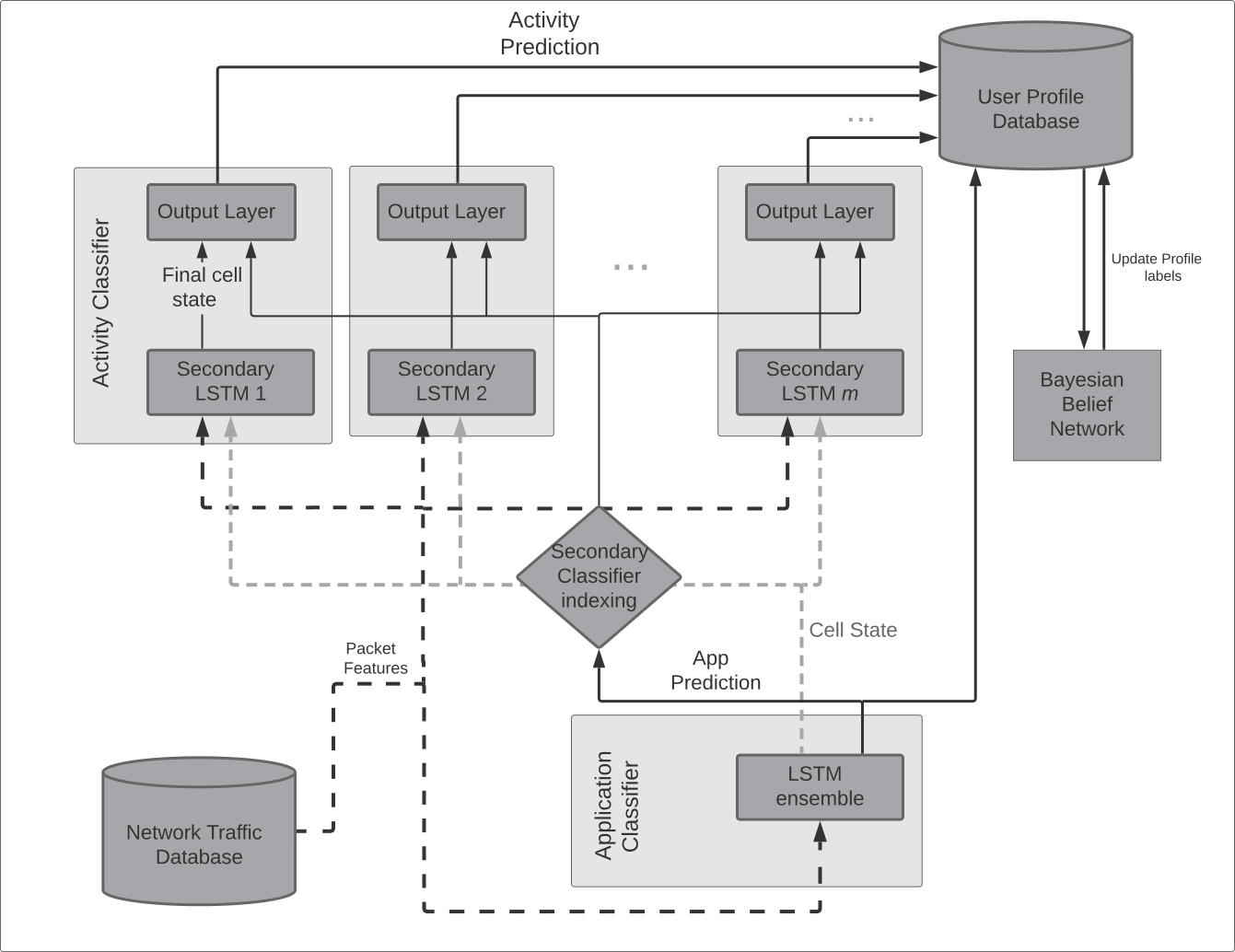}
    \caption{Proposed user profile model (Non-solid lines depict time series dataflow).}
    \label{fig:finalmodel}
\end{figure}

\subsection{User Profiling}
The Bayesian learning approach addresses uncertainty in stochastic processes by treating model parameters as random variables rather than known measurable quantities. It allows us to look for trends in large scale control systems such as producer-consumer models that do not intuitively capture expert attention. The Bayesian Belief Network (BBN) algorithm models state variables with probabilistic estimators and measure the most likely classification using network weights based on observation tables. When appropriately designed to represent real life variables, BBNs are very powerful and under bounded problem scales, na\"ive Bayes classifiers, which assume posterior probability for a class as independent of effects of model attributes on other classes, also offer a practical solution in the absence of expert knowledge of process schematics if the observation labels are duly chosen.\\ Fig. \ref{fig:finalmodel} shows the proposed layout to leverage user history for generating profile characteristics. A user profile database updates its statistics each time a labeled stream for a targeted user is made available, although the streams used to generate profiles might be selected and clustered algorithmically to maximise confidence. These long-term statistics can then periodically drive a Bayesian classifier to track user trait labels.  

\section{Preliminary Results}
\subsection{Experimental setup} \label{fx}
We devise a testing framework for comparing our proposed model against state of the art methods. The time series for every stream is essentially two incoming and outgoing halves. For these three time series, including the original complete time series, mean, standard deviation, minimum, maximum, and similar statistics for packet sizes are used as features for non time series model. Apart from this, we also keep track of the number of packets as well as the sums of squares of the time series as this allows us to calculate combined mean and standard deviation for two merged time series. Such a combination enables training a module that predicts application class for a \textit{cluster} of streams. These probabilistic labels boost the profiling classifier. 



For comparision of proposed model, we have considered random forest of 50 estimators with a depth (maximum number of splits) set to 15 as an \textit{application-classifier} and forests of 20 estimators and maximum depth of 10 as \textit{activity-classifiers} in our comparison. Random forests and SVM models were implemented using  \texttt{RandomForestClassifier} and \texttt{SVC} classes in the scikit-learn library for data analysis and machine learning \cite{sklearn}.

\subsection{Observations}
\begin{figure}[t]
    \centering
    \includegraphics[width=\columnwidth]{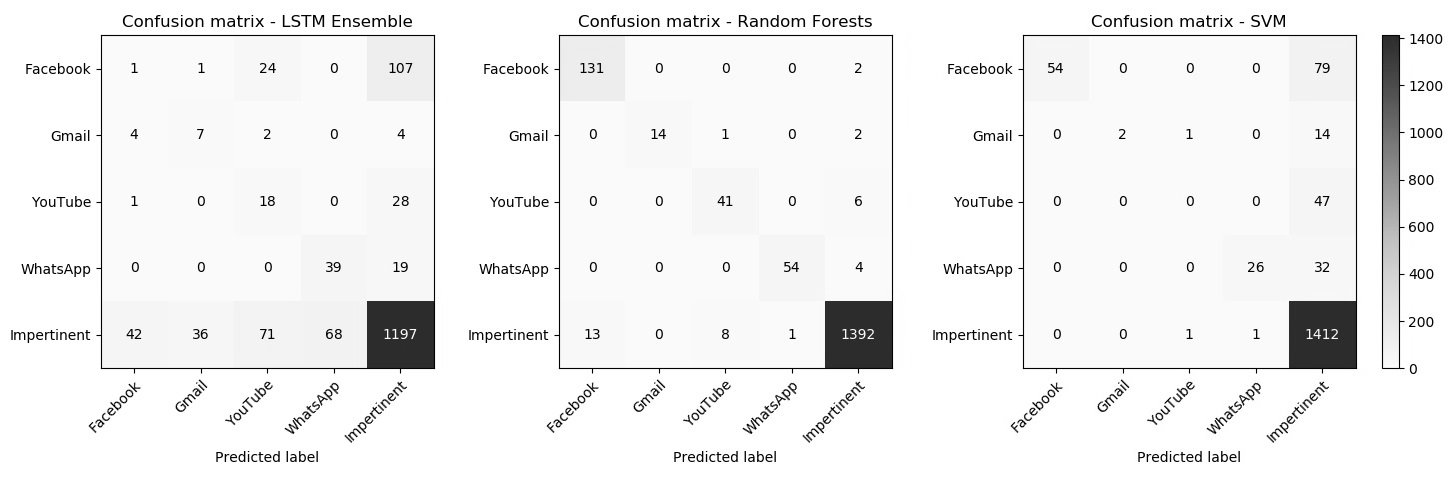}
    \caption{Confusion matrices for the models with 20\% of the preliminary dataset as the testing data. Random forests fared the best on the limited data available.}
    \label{fig:cm}
\end{figure}


Preliminary results over samples with added noise are shown as a means to probe initial behaviour of the models on the data format. The recurrent model was trained with 2000 batches of 50 samples, by which point further training largely converges on the small dataset. For the deep model, streams were sampled uniformly across classes and randomly within a class as uniform sampling across the dataset causes the model to underfit. Accuracy on noised samples of the dataset is shown in Table \ref{accuracy}. Fig. \ref{fig:cm} shows confusion matrices. The proposed method is found to have 87\% accuracy for application classification which essentially means that the system can accurately identify whether the traffic belongs to facebook, youtube or whatsapp. It can also be observed that the proposed method outperforms for the activity classification over YouTube and WhatsApp data. However, its performance is found to be similar to random-forest and SVM for facebook application.

The neural network is heavily dependent on weights set during initialisation and to the number of training epochs where accuracy often drops by the end of the training phase; optimal training requires multiple tests. Contrary to SVMs the LSTM model underfits the impertinent class, suggesting that the model might perform better when sampling is partially biased to dataset proportions. These results however significantly motivate increasing the size of the dataset as all models were clearly able to interpret variations in their feature space. 

\begin{table}[t]
\centering
\caption{Accuracies of application classifier, and activity classifiers for Facebook, YouTube and WhatsApp.}
\begin{tabular}{|c|c|c|c|c|}

\hline
Methodology    & Application & Facebook & YouTube & WhatsApp \\
\hline
Random forests & 0.91 & 0.83     & 0.82    & 0.69     \\

SVM            & 0.84 & 0.64     & 0.55    & 0.59    \\

\textbf{Ensemble LSTM}  & 0.87 & 0.70     & \textbf{0.85}    & \textbf{0.71}     \\
\hline
\end{tabular}
\label{accuracy}
\end{table}

\section{Conclusions}
In this paper, we presented a learning framework to identify the application and the exact activity performed by a smartphone use, using the encrypted traffic generated by smartphone applications. Further, our framework uses the identified activities as the input dataset, to train the model which can profile smartphone users by identifying their age, gender and profession. We also, proposed a method for ground truth construction  that can not only be used to train the model for application and activity classification but also to the learning models for user profiling. To the best of our knowledge this is the first attempt to user profiling using encrypted traffic data.

\subsection{Future research}
There is a need to expand the number of samples in the dataset and adding more activities to the dataset. As their might be several streams servicing a single operation, 
a framework such as our could be greatly boosted with the ability to cluster streams based on \textit{continuous activity} and not just application labels to trace a realistic online fingerprint. Once a model has been developed for gauging application usage, it can be used to create training vectors for successor classifiers at runtime, to power online learning of control systems without needing to store client data locally and hence has major applications for quality-of-service provisioning and authoritative monitoring.

\subsection*{Acknowledgement}
This work, CAIR/CARS 46, is supported by Center for Artificial Intelligence and Robotics (CAIR), Lab. of Defence Research and  Development Organization (DRDO), Bangalore India, under the scheme Contract Acquisition of Research Service (CARS).


\end{document}